\documentclass[aps,prc,twocolumn,floatfix]{revtex4-1}
\usepackage{amsmath}
\usepackage[english]{babel}
\usepackage{amssymb}
\usepackage{graphicx}
\usepackage{bm}
\usepackage{hyperref}
\usepackage{dcolumn}
\usepackage{color}

\def\trmin{{\rm tr}}
\def\mf{{\mbox{\tiny MF}}}

\begin{document}

\title{\Large{Pion masses under intense magnetic fields within the NJL model}}

\author{M. Coppola$^{1,2}$}
\author{D. G\'omez Dumm$^{3}$}
\author{N. N. Scoccola$^{1,2,4}$}

\affiliation{$^{1}$ CONICET, Rivadavia 1917, (1033) Buenos Aires, Argentina}
\affiliation{$^{2}$ Physics Department, Comisi\'{o}n Nacional de Energ\'{\i}a At\'{o}mica, }
\affiliation{Av.\ Libertador 8250, (1429) Buenos Aires, Argentina}
\affiliation{$^{3}$ IFLP, CONICET $-$ Departamento de F\'{\i}sica, Fac.\ de Cs.\ Exactas,
Universidad Nacional de La Plata, C.C. 67, (1900) La Plata, Argentina}
\affiliation{$^{4}$ Universidad Favaloro, Sol{\'{\i}}s 453, (1078) Buenos Aires, Argentina}

\begin{abstract}
The behavior of charged and neutral pion masses in the presence of a static uniform
magnetic field is studied in the framework of the two-flavor Nambu-Jona-Lasinio (NJL) model. Analytical
calculations are carried out employing the Ritus eigenfunction method. Numerical results are obtained for definite model parameters, comparing the predictions of the model with present lattice QCD (LQCD) results.
\end{abstract}


\maketitle

\renewcommand{\thefootnote}{\arabic{footnote}}
\setcounter{footnote}{0}


The study of the behavior of strongly interacting matter under intense
external magnetic fields has gained increasing interest in the last few
years, especially due to its applications to the analysis of relativistic
heavy ion collisions and the description of compact objects like
magnetars~\cite{Andersen:2014xxa}. In this work we concentrate on the effect
of an intense external magnetic field on $\pi$ meson properties. This issue
has been studied in the last years following various theoretical approaches
for low-energy QCD, such as NJL-like models, chiral
perturbation theory, path integral Hamiltonians and LQCD
calculations (see e.g. \cite{Andersen:2014xxa} and refs therein). In the
framework of the NJL model, mesons are usually described as quantum
fluctuations in the random phase approximation
(RPA)~\cite{Klevansky:1992qe}. In the presence of a magnetic field, the
corresponding calculations require some special care, due to the appearance
of Schwinger phases~\cite{Schwinger:1951nm} associated with quark
propagators. For the neutral pion these phases cancel out, and as a
consequence the usual momentum basis can be used to diagonalize the
corresponding polarization
function~\cite{Fayazbakhsh:2012vr,Avancini:2015ady,Avancini:2016fgq,Mao:2017wmq}.
On the other hand, for charged pions Schwinger phases do not cancel,
leading to a breakdown of translational invariance that prevents to proceed
as in the neutral case. In this contribution we present a method
based on the Ritus eigenfunction approach~\cite{Ritus:1978cj} to magnetized
relativistic systems, which allows us to fully
diagonalize the charged pion polarization function. Further details of
this work can be found in Ref.~\cite{Coppola:2018vkw}.
\vspace{3mm}

We start by considering the Euclidean Lagrangian density for the NJL
two-flavor model in the presence of an electromagnetic field. One has
\begin{equation}
{\cal L}  =  \bar \psi \left(- i\, \rlap/\!D + m_0 \right) \psi - G \left[
(\bar\psi \, \psi)^2 + (\bar\psi\, i\gamma_{5}\vec{\tau}\,\psi) \right] \ ,
\label{lagrangian}
\end{equation}
where $\psi = (u\ d)^T$, $\tau_i$ are the Pauli matrices, and
$m_0$ is the current quark mass, which is assumed to be equal for
$u$ and $d$ quarks. The interaction between the fermions and the
electromagnetic field ${\cal A}_\mu$ is driven by the covariant
derivative $D_\mu =  \partial_{\mu}-i\,\hat Q \mathcal{A}_{\mu}$
where $\hat Q=\mbox{diag}(q_u,q_d)$, with $q_u=2e/3$ and $q_d =
-e/3$, $e$ being the proton electric charge. We consider here an
homogeneous stationary magnetic field along the 3 axis in the
Landau gauge, $\mathcal{A}_\mu = B\, x_1\, \delta_{\mu 2}$.

To study meson properties it is convenient to introduce scalar and
pseudoscalar fields $\sigma(x)$ and $\vec{\pi}(x)$, integrating out the
fermion fields. The bosonized Euclidean action is given
by~\cite{Klevansky:1992qe}
\begin{equation}
S_{\mathrm{bos}} \!=\!  -\log\det\mathcal{D}+\frac{1}{4G} \!
\int \! d^{4}x
\Big[\sigma(x)\sigma(x)+ \vec{\pi}(x)\vec{\pi}(x)\Big]\ .
\label{sbos}
\end{equation}
We proceed by expanding this effective action in powers of the fluctuations
$\delta\sigma(x)$ and $\delta\pi_i(x)$ around the corresponding mean field
(MF) values. As usual, we assume that the field $\sigma(x)$ has a nontrivial
translational invariant MF value $\bar{\sigma}$, while the vacuum
expectation values of pseudoscalar fields are zero. In this way one has
\begin{eqnarray}
S_{\mathrm{bos}} \ = \ S^{\mbox{\tiny MF}}_{\mathrm{bos}} \, + \,
S^{\mbox{\tiny quad}}_{\mathrm{bos}}\, + \,\dots
\end{eqnarray}
Here, the mean field action per unit volume reads
\begin{equation}
\frac{S^{\mbox{\tiny MF}}_{\mathrm{bos}}}{V^{(4)}}  =  \frac{ \bar
\sigma^2}{4 G} - \frac{N_c}{V^{(4)}} \!\! \sum_{f=u,d} \int d^4x \, d^4x' \
\trmin\, \ln \left(\mathcal{S}^{\mbox{\tiny MF},f}_{x,x'}\right)^{-1}\!\!\!,
\label{seff}
\end{equation}
where $\trmin$ stands for the trace in Dirac space. The quadratic contribution
can be written as
\begin{equation}
S^{\mbox{\tiny quad}}_{\mathrm{bos}} = \dfrac{1}{2} \!
\sum_{M=\sigma,\pi^r} \int d^4x \, d^4x'\, \delta M
(x)^\ast G_M(x,x')\, \delta M(x')\ , \label{actionquad}
\end{equation}
where $r=0,\pm$ with $\pi^\pm=\left(\pi_1 \mp i \pi_2\right)/\sqrt{2}$, and
\begin{eqnarray}
G_M(x,x')&=&\frac{1}{2 G}\; \delta^{(4)}(x-x') - J_M(x,x')\ , \nonumber \\
J_{\pi^0} (x,x') &=& N_c \sum_f \trmin \bigg[
\mathcal{S}^{\mf,f}_{x,x'} \ \gamma_5 \ \mathcal{S}^{\mf,f}_{x',x}
\ \gamma_5 \ \bigg]\ ,
\nonumber \\
J_{\pi^\pm} (x,x') &=& 2 N_c  \, \trmin \bigg[
\mathcal{S}^{\mf,u}_{x,x'} \ \gamma_5 \ \mathcal{S}^{\mf,d}_{x',x}
\ \gamma_5 \ \bigg]\ . \label{jotas}
\end{eqnarray}
The expression for $J_\sigma$ is obtained from $J_{\pi^0}$ replacing
$\gamma_5$ matrices with unit matrices in Dirac space. In these expressions we have introduced the mean field
quark propagators $\mathcal{S}^{\mf,f}_{x,x'}$.
As is well known, their explicit form can be written in different
ways~\cite{Andersen:2014xxa}. For convenience we take here a form given by a
product of a phase factor and a translational invariant function, namely
\begin{equation}
S^{\mf,f}_{x,x'} \ = \ e^{i\Phi_f(x,x')}\,\int \dfrac{d^4 p}{(2\pi)^4}\ e^{i\, p\, (x-x')}\, \tilde S_p^f\
,
\label{sfx}
\end{equation}
where $\Phi_f(x,x')=\exp\big[i q_f B (x_1+x_1')(x_2-x_2')/2 \big]$ is the
so-called Schwinger phase.
We express now $\tilde S_p^f$ in the Schwinger form~\cite{Andersen:2014xxa}
\begin{eqnarray}
&& \tilde S_p^f = \int_0^\infty \! d\tau\,
\exp\!\bigg[-\tau\Big(M^2+p_\parallel^2+p_\perp^2\,\dfrac{\tanh\tau B_f}{\tau B_f}\Big) \bigg] \times \nonumber\\
&& \left[\left(M\!-\!p_\parallel \gamma_\parallel\right)\,\left(1\!+\!i s_f \gamma_1 \gamma_2 \tanh\tau B_f\right) \!-\!
\dfrac{p_\perp \gamma_\perp}{\cosh^2 \tau B_f}  \right]\ ,
\label{sfp_schw}
\end{eqnarray}
where we have introduced some definitions. The perpendicular and parallel
gamma matrices are collected in vectors $\gamma_\perp = (\gamma_1,\gamma_2)$
and $\gamma_\parallel = (\gamma_3,\gamma_4)$. Similarly,
$p_\perp = (p_1,p_2)$ and $p_\parallel = (p_3,p_4)$. The quark
effective mass $M$ is given by $M=m_0+\bar\sigma$, while
$s_f = {\rm sign} (q_f B)$ and $B_f=|q_fB|$. Notice that the integral in
Eq.~(\ref{sfp_schw}) is divergent and has to be properly regularized, as we
discuss below.

\hfill

At the MF level, one arrives to the usual gap equation by replacing in
Eq.~(\ref{seff}) the above expression for the quark propagator and
minimizing with respect to $M$. It can be seen that if we regularize this
equation using the Magnetic Field Independent Regularization (MFIR)
scheme~\cite{Menezes:2008qt,Allen:2015paa} together with a 3D cutoff, the
resulting expression is in
agreement with the corresponding one given in Ref.~\cite{Klevansky:1992qe}.
Moreover, it also matches the result obtained in Ref.~\cite{Menezes:2008qt},
where the propagator is expressed in terms of a sum over Landau levels.

\hfill

As for the pion masses, we notice that the analysis of the $\pi^0$ pole mass
in the presence of a magnetic field within the MFIR scheme has already been
carried out in Refs.~\cite{Avancini:2015ady,Avancini:2016fgq}. However, in
those works the authors use a representation of the quark propagator
different from the Schwinger one in Eqs.~(\ref{sfx}-\ref{sfp_schw}). Thus,
we find it opportune to verify that both representations lead to the same
results for the $\pi^0$ mass. We start by replacing Eq.~(\ref{sfx}) into the
expression for the polarization function $J_{\pi^0}(x,x')$ in
Eq.~(\ref{jotas}). The contributions of the Schwinger phases to each term of
the sum correspond to the same quark flavor, hence, they cancel out. As a
consequence, the polarization function depends only on the difference $x-x'$
(i.e., it is translational invariant), which leads to the conservation of
$\pi^0$ momentum. If we take now the Fourier transform of the $\pi^0$ fields
to the momentum basis, the corresponding transform of the polarization
function will be diagonal in $q,q'$ momentum space. Thus, the $\pi^0$
contribution to the quadratic action in the momentum basis can be written as
\begin{equation}
S^{\mbox{\tiny quad}}_{\pi^0}  =  \dfrac{1}{2} \int \dfrac{d^4 q}{(2\pi)^4}\ \
\delta\pi^0(-q) \, \left[\frac{1}{2 G} -
J_{\pi^0}(q_\perp^2,q_\parallel^2) \right] \delta\pi^0(q)\ .
\label{actionquadpi0p}
\end{equation}
Choosing the frame in which the $\pi^0$ meson is
at rest, its mass can be obtained by solving the equation
\begin{equation}
\frac{1}{2 G} - J^{\rm (reg)}_{\pi^0}(0,-m_{\pi^0}^2) \ = \ 0\ ,
\end{equation}
where $J^{\rm (reg)}_{\pi^0}(0,-m_{\pi^0}^2)$ is obtained from
$J_{\pi^0}(0,-m_{\pi^0}^2)$ after some regularization procedure. Using the
MFIR scheme, it can be shown that ---as in the case of the gap equation---
our result for $J^{\rm(reg)}_{\pi^0}(0,-m_{\pi^0}^2)$ agrees with the
corresponding expression obtained in Ref.~\cite{Avancini:2015ady}, where the
calculation has been done using an expansion in Landau levels for the quark
propagators instead of considering the Schwinger form in
Eq.~(\ref{sfp_schw}).

\hfill

Let us focus on the study of charged pion masses. We will consider the
$\pi^+$ meson, although a similar analysis can be carried out for the
$\pi^-$, leading to the same expression for the $B$-dependent mass. Once again,
we replace Eq.~(\ref{sfx}) into the expression for the polarization function
$J_{\pi^+}(x,x')$ in Eq.~(\ref{jotas}). Now, in contrast to the $\pi^0$ case, it is
seen that the Schwinger phases do not cancel, due to their different quark
flavors. Therefore, the $\pi^+$ polarization function is not translational
invariant, and consequently it will not become diagonal when transformed to
the momentum basis. In this situation we find it convenient to follow the
Ritus eigenfunction method~\cite{Ritus:1978cj}. Namely, we expand the
charged pion field as
\begin{equation}
\pi^+(x) \ = \ \dfrac{1}{2\pi}\sum_{k=0}^\infty
\left[ \prod_{i=2}^4 \int \frac{dq_i}{2\pi}\right]
\mathbb{F}_{\bar q}^+(x) \,
\pi_{\bar q}^+ \ ,
\label{Ritus}
\end{equation}
where ${\bar q}=(k,q_2,q_3,q_4)$ and
\begin{equation}
\mathbb{F}_{\bar q}^+(x) \ = \ N_k \, e^{i ( q_2 x_2 + q_3 x_3 + q_4 x_4)}
\, D_k(\rho_{+})\ .
\label{Fq}
\end{equation}
Here $D_k(x)$ are the cylindrical parabolic functions, and we have used the
definitions $N_k= (4\pi B_{\pi^+})^{1/4}/\sqrt{k!}$ and $\rho_+ =
\sqrt{2B_{\pi^+}}\,x_1-s_+\sqrt{2/B_{\pi^+}}\,q_2$, where $B_{\pi^+} =
|q_{\pi^+} B|$ and $s_+= \mathrm{sign}(q_{\pi^+} B)$, with $q_{\pi^+} = q_u
- q_d = e$. In this basis the charged pion polarization function becomes diagonal. The corresponding contribution to the quadratic action in Eq.~(\ref{actionquad}) is given by
\begin{equation}
S^{\mbox{\tiny quad}}_{\pi^+} \! = \! \dfrac{1}{4\pi}\sum_{k=0}^\infty
\left[\prod_{i=2}^4 \int \! \frac{dq_i}{2\pi}\right]
\!(\delta\pi_{\bar q}^+)^\ast \!\!
\left[ \frac{1}{2G} - J_{\pi^+}(k,\Pi^2)\right]\!
\delta\pi_{\bar q}^+,
\label{actionquadTPF}
\end{equation}
where $\Pi^2=(2k+1)\, B_{\pi^+}+q_\parallel^2$ and
\begin{eqnarray}
&& J_{\pi^+}(k,\Pi^2) = \dfrac{N_c}{2\pi^2} \int_0^\infty\! dz
\int_{0}^1 dy \ \frac{e^{-zy(1-y)\left[\Pi^2-(2k+1)B_{\pi^+} \right]}}{\alpha_+} \nonumber \\
&& e^{-zM^2} \left(\dfrac{\alpha_-}{\alpha_+}\right)^k \Bigg\{
\dfrac{(1-t_u^2)(1-t_d^2)}{\alpha_+\,\alpha_-}\, \Big[ \alpha_- + (\alpha_- - \alpha_+)\,k \Big] + \nonumber \\
&& \left[M^2+\dfrac{1}{z}-y(1-y)\left(\Pi^2-(2k+1)\,
B_{\pi^+} \right) \right](1-t_u \,t_d) \Bigg\}\ . \nonumber\\
\label{J+B}
\end{eqnarray}
Here we have defined $t_u=\tanh (B_u y z)$, $t_d=\tanh [B_d (1-y) z]$ and
$\alpha_\pm = (B_d t_u+B_u t_d \pm
 B_{\pi^+} \,t_ut_d)/(B_u B_d)$.

Once again, we carry out a regularization within the MFIR scheme,
using a 3D cutoff. We obtain
\begin{equation}
J_{\pi^+}^{{\rm (reg)}}(k,\Pi^2) \ = \ \, J^{\rm
(reg)}_{\pi,B=0}(\Pi^2) \, +  \, J_{\pi^+}^{{\rm (mag)}}(k,\Pi^2)\
, \label{J+reg}
\end{equation}
where $J_{\pi^+}^{{\rm (mag)}}(k,\Pi^2)$ is finite and $J^{\rm (reg)}_{\pi,B=0}(\Pi^2)$ corresponds
to the usual pion polarization function in the absence of magnetic field evaluated
at $q^2 = \Pi^2$. It can be easily seen
that the same polarization function is obtained for the case of the $\pi^-$ meson.

For a point-like pion in Euclidean space, the two-point function will vanish when $\Pi^2 = - m_{\pi^+}^2$ or, equivalently, $q_\parallel^2 = - [ m_{\pi^+}^2 + (2k+1)\,eB]$, for a
given value of $k$. Therefore, in our framework the
charged pion pole mass can be obtained for each Landau level $k$ by solving the
equation
\begin{equation}
\frac{1}{2G} - J_{\pi^+}^{{\rm (reg)}}(k,-m_{\pi^+}^2) \ = \ 0 \ .
\end{equation}
Of course, while for a point-like pion $m_{\pi^+}$ is a B-independent quantity (the $\pi^+$ mass in vacuum), in the
present model ---which takes into account the internal quark structure of the pion--- it depends on the magnetic field. Instead of dealing with this quantity, it has become customary in the literature to define the $\pi^+$
``magnetic field-dependent mass'' (MFDM) as the lowest
quantum-mechanically allowed energy of the $\pi^+$ meson, namely
\begin{equation}
E_{\pi^+}(eB) \! = \!  \sqrt{m_{\pi^+}^2 + \Pi^2 - q_3^2}\, \bigg|_{\substack{
q_3 =0 \\ k=0}}
\! = \! \sqrt{m_{\pi^+}^2 + eB}\: ,
\label{epimas}
\end{equation}
(see e.g.~Ref.~\cite{Bali:2017ian}). Notice that this ``mass'' is magnetic
field-dependent even for a point-like particle. In fact, owing to zero-point
motion in the 1-2 plane, even for $k=0$ the charged pion cannot be at rest
in the presence of the magnetic field.

\hfill

To get numerical predictions we consider some model parameterizations that
reproduce not only low-energy phenomenological vacuum properties but also
LQCD results for the behavior of quark-antiquark condensates under an
external magnetic field. Let us consider the parameter set $m_0 = 5.66$ MeV,
$\Lambda = 613.4$ MeV and $G\Lambda^2 = 2.250$, which (for vanishing
external field) corresponds to an effective mass $M=350$~MeV and a
quark-antiquark condensate $\langle \bar f f\rangle (B=0) = (-243.3\ {\rm
MeV})^3$. We denote this parameterization as Set I. To test the sensitivity
of our results with respect to the model parameters we will consider two
alternative parameterizations, denoted as Set II and Set III, which
correspond to $M=320$ and 380~MeV, respectively. All these parameter sets
properly reproduce the empirical values of the pion mass and decay constant
in vacuum, $m_\pi=138$~MeV and $f_\pi=92.4$~MeV. As discussed in
Ref.~\cite{Coppola:2018vkw}, they also provide a very good agreement with
the lattice results quoted in Ref.~\cite{Bali:2012zg} for the quark
condensates under an external magnetic field. In fact, it is seen that the
predictions are not significantly affected by the parameter choice.

\hfill

\begin{figure}[htb]
\centering{}\includegraphics[width=0.9\columnwidth]{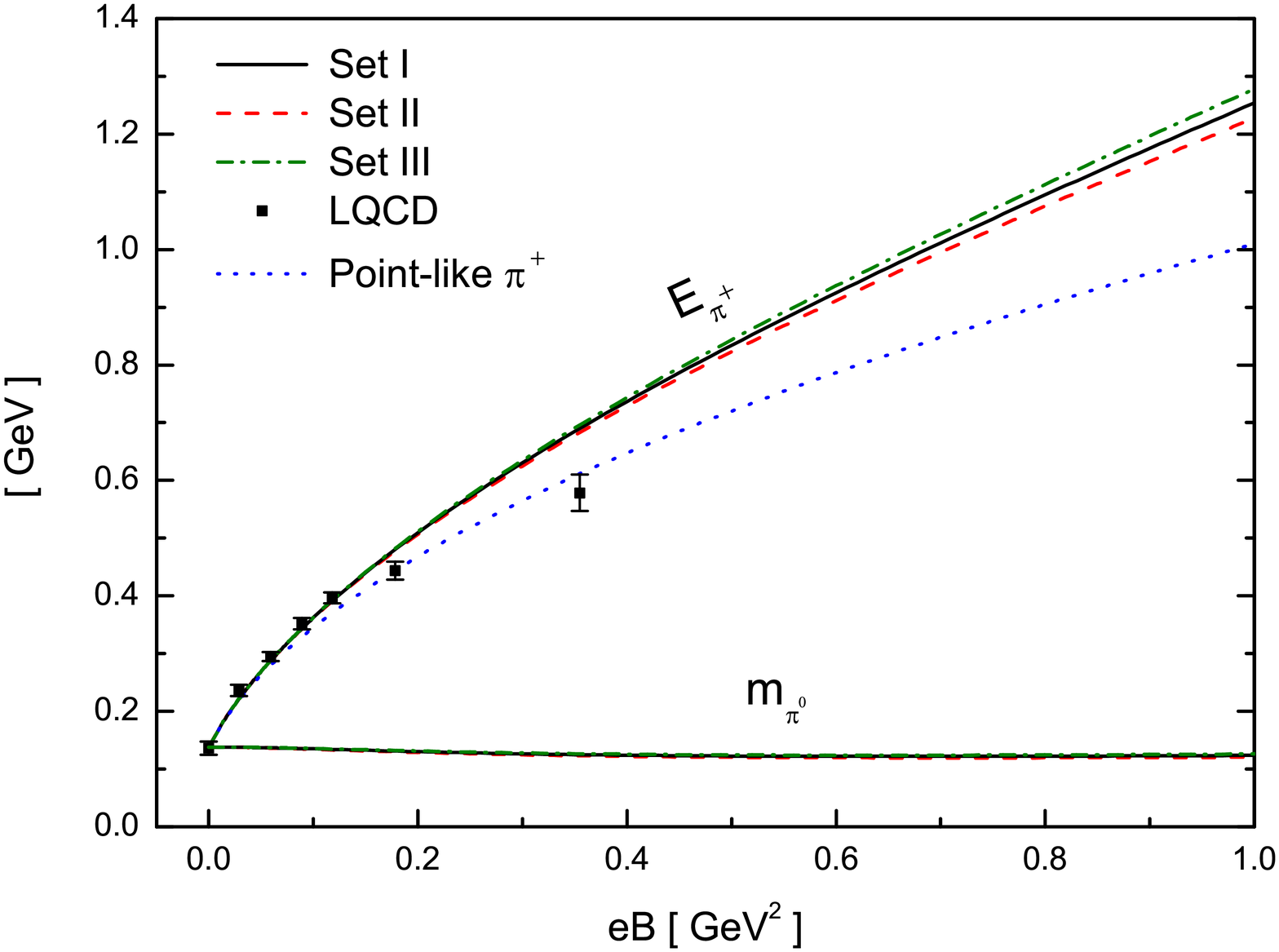} \caption{(Color
online) $\pi^0$ mass and $\pi^+$ MFDM as functions of $eB$ for different model parameter sets.
The MDFM  of a point-like $\pi^+$ pion (dotted line) as well as results from LQCD
calculations in Ref.~\cite{Bali:2011qj} (squares) are included for comparison.} \label{fig1}
\end{figure}

In Fig.~\ref{fig1} we show our numerical results for the behavior of pion
masses, which are plotted as functions of $eB$.  In the case of the $\pi^+$,
the curves correspond to the MFDM defined by Eq.~(\ref{epimas}). As can be
seen, the model predicts an increasing enhancement of $E_{\pi^+}$ with the
magnetic field. For comparison, we also show the behavior of $E_{\pi^+}$ for
the case of a point-like meson and the LQCD results quoted in
Ref.~\cite{Bali:2011qj}. These LQCD calculations consider realistic pion
masses and values of $eB$ up to $\sim 0.4$~GeV$^2$, using staggered quarks.
It is found that model predictions are in good agreement with LQCD results
for $eB\lesssim 0.15$ GeV$^2$, while they seem to deviate from them for
larger values of the magnetic field. Concerning the $\pi^0$ mass, it is seen
that it shows a slight decrease with $eB$, as previously found e.g. in
Refs.~\cite{Avancini:2015ady,Avancini:2016fgq}. Once again the results are
in general rather independent of the model parametrization.

Besides the mentioned LQCD calculation in Ref.~\cite{Bali:2011qj}, more
recent lattice simulations using Wilson
fermions~\cite{Bali:2017ian,Bali:2017yku} have been carried out, providing
results for $\pi^+$ and $\pi^0$ masses for larger values of $eB$. In these
simulations, however, a heavy pion with $m_\pi(0)=415$~MeV in vacuum has
been considered. In order to compare these results with our predictions we
follow the procedure carried out in Ref.~\cite{Avancini:2016fgq}, viz.\ we
consider a new parameter Set Ib in which $G$ and $\Lambda$ are the same as
in Set I, while $m_0$ is increased so as to obtain $m_\pi(0) = 415$~MeV. In
Ref.~\cite{Avancini:2016fgq} the authors also consider a magnetic field
dependent coupling of the form $G(eB)= \alpha+\beta\,\exp[-\gamma\,(eB)^2]$
in order to reproduce LQCD results for both the behavior of quark
condensates and the $\pi^0$ mass.

\begin{figure}[h]
\centering{}\includegraphics[width=0.9\columnwidth]{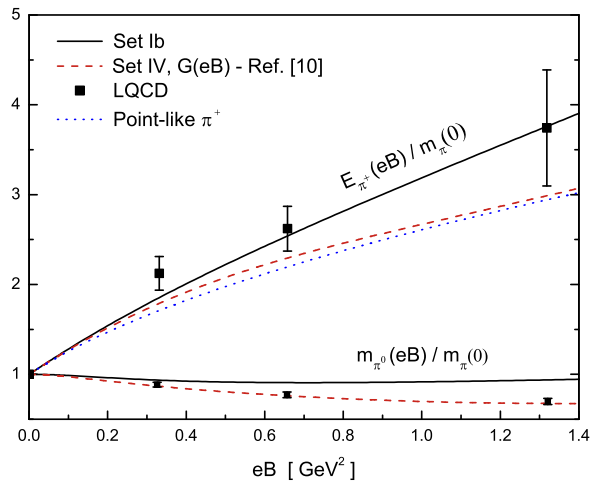} \caption{(Color
online) Normalized $\pi^0$ mass and $\pi^+$ MFDM as functions of $eB$ for Set Ib (solid lines)
and Set IV of Ref.~\cite{Avancini:2016fgq} (dashed lines). The normalized MFDM of a point-like $\pi^+$
(dotted line) and results from LQCD simulations in Ref.~\cite{Bali:2017ian} (squares),
which consider a $B=0$ pion mass of 415~MeV, are included for comparison.}
\label{fig2}
\end{figure}

The curves for the normalized charged pion MFDM, $E_{\pi^+}/m_\pi(0)$,
and the neutral pion mass, $m_{\pi^0}/m_\pi(0)$, for Set Ib are shown
in Fig.~\ref{fig2}, together with LQCD results obtained for these quantities
after an extrapolation of lattice spacing to the continuum~\cite{Bali:2017ian}.
Results corresponding to the parameter Set IV of Ref.~\cite{Avancini:2016fgq},
with the $B$-dependent coupling $G(eB)$, are also included. It is seen that for
the $\pi^+$ meson the results from Set Ib are consistent with lattice data,
although the errors in the latter are considerably large to be conclusive (in
fact, results obtained considering finite lattice spacings become closer to the
point-like $\pi^+$ curve~\cite{Bali:2017yku}). On the other hand, in the case of
the $\pi^0$ mass, where errors from LQCD are smaller, the curve obtained from Set Ib
lies above lattice predictions. Regarding the model proposed in
Ref.~\cite{Avancini:2016fgq}, it is seen that the behavior of the $\pi^+$ normalized
MFDM is similar to that of a point-like particle, while (as discussed in
Ref.~\cite{Avancini:2016fgq}) the results for the $\pi^0$ mass are in good agreement
with LQCD data. It is worth noticing that, in that model, once $m_0$ is rescaled to get
a phenomenologically acceptable value for the pion mass, the corresponding parametrization
leads to a too low value for the pion decay constant at $B=0$ of $f_\pi\simeq 80$ MeV.

\hfill

In conclusion, we have analyzed the effect of an intense homogeneous external magnetic
field on $\pi$ meson masses within the two-flavor NJL model. In
particular, we have shown that the Ritus eigenfunction method diagonalizes the charged pion polarization function, fully taking into account the translational-breaking effects introduced by the Schwinger phases in the RPA approach.

In our numerical calculations we have used different model parameterizations
that satisfactorily describe not only meson properties in the absence of the
magnetic field but also the behavior of quark condensates as functions of
$B$ obtained in LQCD calculations.  We have found that when the magnetic
field is enhanced, the $\pi^0$ mass shows a slight decrease, while the MFDM
of the charged pion steadily increases, remaining always larger than that of
a point-like pion. These results are in agreement with LQCD calculations
with realistic pion masses for low values of $eB$ (say $eB\lesssim 0.15$
GeV$^2$), although there seems to be some discrepancy as the magnetic field
is increased. For larger values of $eB$, some recent LQCD simulations for
$m_{\pi^0}$ and $E_{\pi^+}$ have been carried out considering unphysically
large quark masses. In the case of $E_{\pi^+}$ the results are consistent
with our calculations (with adequately rescaled parameters), while there is
a significant discrepancy in the case of the $\pi^0$ mass. The agreement for
$m_{\pi^0}$ gets improved if, as done in Ref.~\cite{Avancini:2016fgq}, a
magnetic field-dependent coupling $G(eB)$ is introduced. In this sense, it is
worth noticing that nonlocal NJL-like models, which naturally predict a magnetic
field dependence of the quark current-current interaction, are also able to
reproduce adequately the $\pi^0$ mass behavior~\cite{GomezDumm:2017jij}.

\hfill

This work has been
supported in part by CONICET and ANPCyT (Argentina), under grants PIP14-578,
PIP12-449, and PICT14-03-0492, and by the National University of La Plata
(Argentina), Project No.\ X718.

\end{document}